\documentclass[aps,pre,showpacs,twocolumn,amsmath,amssymb,superscriptaddress,floatfix]{revtex4-2} 
\usepackage{graphicx}
\usepackage{dcolumn}
\usepackage{bm}

\newcommand{\gen}{n}
\newcommand{\G}{\mathcal{G}}
\newcommand{\GR}{\mathcal{G}_{\text{R}}}
\newcommand{\Ggen}{G_{\text{gen}}}
\newcommand{\mgen}{m_{\text{gen}}}
\newcommand{\mgenav}{\langle m_{\text{gen}}\rangle}
\newcommand{\mgeni}{m_{\text{gen}}^{(i)}}

\newcommand{\kappaav}{\langle \kappa\rangle}
\newcommand{\lambdaeff}{\langle\!\langle \lambda \rangle\!\rangle}
\newcommand{\Df}{D_{\text{f}}}
\newcommand{\df}{d_{\text{f}}}
\newcommand{\dfmin}{d_{\text{f}}^{\text{min}}}
\newcommand{\dfmax}{d_{\text{f}}^{\text{max}}}
\newcommand{\ds}{d_{\text{s}}}
\newcommand{\Ds}{D_{\text{s}}}
\newcommand{\dw}{d_{\text{w}}}
\newcommand{\dk}{d_{\text{k}}}
\newcommand{\dsmin}{d_{\text{s}}^{\text{min}}}
\newcommand{\dsmax}{d_{\text{s}}^{\text{max}}}
\newcommand{\dshub}{d_{\text{s}}^{\text{(hub)}}}
\newcommand{\lB}{l_{\text{B}}}
\newcommand{\NB}{N_{\text{B}}}
\newcommand{\TR}{T_{\text{R}}}
\newcommand{\TB}{T_{\text{B}}}
\newcommand{\Rroot}{R_{\text{root}}}
\newcommand{\Rrooteff}{\langle\!\langle R_{\text{root}} \rangle\!\rangle}

\begin{document}

\preprint{APS/123-QED}

\title{Random walks on bifractal networks}

\author{Kousuke Yakubo}%
\email{yakubo@eng.hokudai.ac.jp}
\affiliation{%
Department of Applied Physics, Hokkaido University, Sapporo 060-8628, Japan}
\author{Gentaro Shimojo}%
\email{sswb\_sswb\_m@eis.hokudai.ac.jp}
\affiliation{%
Department of Applied Physics, Hokkaido University, Sapporo 060-8628, Japan}
\author{Jun Yamamoto}
\email{jun.j.yamamoto@gmail.com}
\affiliation{%
Department of Network and Data Science, Central European University, A-1100 Wien, Austria}
\date{\today}

\begin{abstract}
It has recently been shown that networks possessing scale-free and fractal properties may
exhibit a bifractal nature, in which local structures are described by two different fractal
dimensions. In this study, we investigate random walks on such fractal scale-free networks (FSFNs)
by examining the walk dimension $\dw$ and the spectral dimension $\ds$, to understand how the
bifractality affects their dynamical properties. The walk dimension is found to be unaffected by
the difference in local fractality of an FSFN and remains constant regardless of the starting node
of a random walk, whereas the spectral dimension takes two values, $\dsmin$ and $\dsmax(> \dsmin)$,
depending on the starting node. The dimension $\dsmin$ characterizes the return probability of a
random walker starting from an infinite-degree hub node in the thermodynamic limit, while $\dsmax$
describes that of a random walker starting from a finite-degree non-hub node infinitely distant
from hub nodes and is equal to the global spectral dimension $\Ds$. The existence of two local
spectral dimensions is a direct consequence of the bifractality of the FSFN. Furthermore,
analytical expressions of $\dw$, $\dsmin$, and $\dsmax$ are presented for FSFNs formed by the
generator model and the giant components of critical scale-free random graphs, and are numerically
confirmed.
\end{abstract}


\maketitle

\section{Introduction}
\label{sec:intro}
Multifractality is often exhibited by physical quantities (measures) that are spatially or
temporally distributed in heterogeneous manners with long-range correlations
\cite{Paladin87,Stanley88,Salat17}. For such a multifractal measure, the coarse-grained quantity
$\mu_{b(l)}$, defined as the sum of the measure within a box $b$ of linear size $l$ located at a
specific position, is proportional to $l^{\alpha}$, where the H\"older exponent $\alpha$ depends on
the position of the box. Furthermore, the fractal dimension $f(\alpha)$ of the set of boxes
characterized by $\alpha$ varies with $\alpha$. The multifractal nature has been observed in a
variety of systems, including turbulent flows \cite{Mandelbrot74,Benzi84,Meneveau87,Sreenivasan91},
diffusion-limited aggregation \cite{Amitrano86}, disordered quantum systems
\cite{Aoki83,Schreiber91}, biological systems \cite{Smith96,Lopes09}, and financial markets
\cite{Jiang19}. The concept of multifractality can be applied to topological structures of networks
composed of nodes and edges by replacing the Euclidean distance in its definition with the
shortest-path distance between nodes \cite{Furuya11,Wang12}. Indeed, multifractal measures on
various networks have been extensively studied so far \cite{Rosenberg20,Li14,Liu15,Rendon17,
Pavon20,Xiao21, Ding21,Pavon22,Zhao23,Yamamoto23}. While the H\"older exponent usually takes
continuous values within a certain range, sometimes it takes only two values, as seen in the
scaling behavior of L\'{e}vy flights \cite{Jaffard99,Nakao00,Chechkin00}. The multifractal property
of such systems is specifically referred to as \textit{bifractality}.

The bifractal nature of heterogeneous systems has been studied also for networks that possess
scale-free and fractal properties \cite{Furuya11}. While the term ``scale free" typically refers to
the property that the degree distribution $P(k)$ takes the form $P(k)\propto k^{-\gamma}$ with
$2<\gamma <3$ for high degree $k$, we consider, in this paper, a network scale free if $P(k)$ obeys
a power-law with $\gamma > 2$. In addition, the fractality of a network implies that when
the network is covered with its subgraphs (boxes) of diameter $l$, the minimum number of covering
boxes decreases proportionally to a power of $l$. Real-world networks often exhibit the scale-free
and fractal properties \cite{Song05}. Such networks are called fractal scale-free networks (FSFNs).
The previous work \cite{Furuya11} has shown that FSFNs may exhibit bifractality even for constant
measures. A multifractal constant measure implies that the structure of the underlying support
itself is multifractal, which has already been found in conventional non-network systems
\cite{Vicsek90,Cheng95,Jestczemski96}. Since the H\"older exponent $\alpha$ of a structurally
multifractal system is nothing but the fractal dimension of a local structure, the bifractal
property of an FSFN means that the FSFN is characterized by two local fractal dimensions. Our
recent work has revealed that a wide range of FSFNs are actually bifractal, and it has been
conjectured that any FSFN possesses a bifractal structure \cite{Yamamoto23}.

It is natural to ask how the bifractality impacts dynamical processes on networks. In the absence
of previous works on network dynamics that take into account structural bifractality, it is
significant to elucidate the influence of the structural bifractality on the dynamical properties of
networks. In this work, we focus on random walks on FSFNs to clarify this issue. Random walks on
networks have been extensively studied as fundamental dynamics on complex networks because they are
closely related to diverse phenomena and applications, including information diffusion, infection
spread, cascading failures, and the PageRank algorithm \cite{Burioni05,Masuda17,Riascos21}. Here,
we examine, in particular, the walk and spectral dimensions of random walks on bifractal networks.
The walk dimension determines the dependence of the first passage time on internode distance or
the time dependence of the mean topological displacement, while the spectral dimension describes
the time dependence of the return probability of a random walker returning to the starting node
after $t$ steps. Our primary finding is that the spectral dimension takes two different values,
$\dsmin$ and $\dsmax$, depending on whether the walker starts from a high-degree hub node or from a
non-hub node apart from hubs. This is a direct consequence of the bifractality. In contrast, the
walk dimension $\dw$ remains constant regardless of the starting node of the random walk. Moreover,
analytical expressions of $\dsmin$, $\dsmax$, and $\dw$ are presented for two mathematical models
that generate a wide range of FSFNs.

The rest of this paper is organized as follows: In Sec.~\ref{sec:2}, we argue how the walk and
spectral dimensions are affected by the bifractality of FSFNs by considering the relation between
the first passage time in an FSFN and that in its renormalized network. This leads to the
conclusion that the walk dimension $\dw$ is not affected by the local fractality of the network,
while the spectral dimension takes two distinct values ($\dsmin$ and $\dsmax$) depending on the
starting node of a random walk. Section \ref{sec:3} presents the analytical expressions of $\dw$,
$\dsmin$, and $\dsmax$ for a wide range of FSFNs formed by two types of network models. These
results are numerically confirmed in this section. Finally, in Sec.~\ref{sec:4}, we discuss the
significance of our results in relation to previous studies and conclude our paper.

\section{Random walks on fractal scale-free networks}
\label{sec:2}

Consider a connected and undirected FSFN $\G$ that has a degree distribution with a power-law tail,
$P(k)\propto k^{-\gamma}$ for high degree $k$. Let $\Df$ be the (global) fractal dimension of the
FSFN $\G$. This means that if we cover the network $\G$ with subgraphs (boxes) of diameter $l$, the
minimum number $\NB(l)$ of boxes scales as $\NB(l)\propto l^{-\Df}$ \cite{Song05}. If the number
$\nu_{b}$ of nodes included in a covering box $b$ is proportional to the number $k_{b}$ of
neighboring boxes to $b$, namely,
\begin{equation}
\nu_{b}\propto k_{b} ,
\label{eq:1}
\end{equation}
the network $\G$ exhibits a bifractal nature in its structure \cite{Furuya11}. The local fractal
dimensions of $\G$ are then given by
\begin{subequations}
\begin{align}
\dfmin&= \Df \left(\displaystyle\frac{\gamma - 2}{\gamma - 1}\right) , \label{eq:2a}\\
\dfmax&= \Df . \label{eq:2b}
\end{align}
\label{eq:2}
\end{subequations}
The fact that an FSFN satisfying Eq.~(\ref{eq:1}) is characterized by these two local fractal
dimensions is closely related to the convergence of the $q$-th moment, $Z_{q}(l)=\sum_{b}\mu_{b(l)}^{q}$,
of the box measure $\mu_{b(l)}$. The quantity $Z_{q}(l)$ governs the multifractal property of the
system. If the measure $\mu_{i}$ on node $i$ is independent of $i$ (i.e., constant measure) and
Eq.~(\ref{eq:1}) holds, the box measure $\mu_{b(l)}=\sum_{i\in b(l)}\mu_{i}$ becomes
$\mu_{b(l)}=\nu_{b}/N\propto k_{b}/N$, where $N$ is the number of nodes in $\G$. This relation
allows us to approximate the sum over the boxes in $Z_{q}(l)$ by an integral with respect to $k_{b}$.
The integral converges when $q<\gamma-1$, whereas for $q\ge\gamma-1$ it diverges, necessitating
truncation of the integration range at the natural cut-off $k_{c}(l)\propto l^{-\Df/(\gamma-1)}$.
These two types of convergence of $Z_{q}(l)$ lead to two values of the H\"older exponent, namely
two local fractal dimensions. It has been also clarified that the dimension $\dfmin$ describes the
local fractality around hub nodes with infinitely high degrees in the thermodynamic limit, while
$\dfmax$ represents that around finite-degree non-hub nodes infinitely distant from the hubs
\cite{Yamamoto23}.

In this work, we examine simple random walks on such an FSFN $\G$. The probability $p_{ij}(t)$ that
a random walker starting from node $i$ at $t=0$ is at node $j$ at time $t$ is governed by the
master equation
\begin{equation}
p_{ij}(t)=\sum_{j'}\displaystyle\frac{A_{jj'}}{k_{j'}}p_{ij'}(t-1) ,
\label{eq:3}
\end{equation}
where the adjacency matrix element $A_{ij}$ of $\G$ is $1$ if the nodes $i$ and $j$ are connected
by an edge and $0$ otherwise, and $k_{i}$ is the degree of node $i$.

First, we study the effect of bifractality of $\G$ on the walk dimension $\dw$ which determines the
inter-node distance dependence of the first passage time (FPT). The FPT $T_{ij}$ is the expected
time required for a walker starting from the source node $i$ to reach the target node $j$ for the
first time. In conventional fractal systems or non-scale-free fractal networks of size $N$,
$T_{ij}$ is proportional to $l_{ij}^{\dw}$ for $l_{ij}$ of the order of $N^{1/\Df}$, where $l_{ij}$
is the (shortest-path) distance between nodes $i$ and $j$ \cite{ben-Avraham00}. In the case of a
bifractal network in which the local fractality around a high-degree hub node is different from
that around a low-degree non-hub node, there is a possibility that the $l_{ij}$ dependence of the
FPT, and consequently the walk dimension $\dw$, may depend on $k_{i}$ and $k_{j}$.

In order to clarify this point, we consider the mean time $T(l;k,k')$, which is the average of
$T_{ij}$ over node pairs $(i,j)$ satisfying $k_{i}=k$, $k_{j}=k'$, and $l_{ij}=l$. The scaling
behavior of $T(l;k,k')$ has been thoroughly studied in Ref.~\cite{Gallos07}, where the authors
considered the renormalized network $\GR$ of a given network $\G$ constructed by covering $\G$ with
the minimum number of boxes of linear size $\lB$ and regarding each box as a supernode of $\GR$. In
the renormalized network $\GR$, supernodes corresponding to boxes $b$ and $b'$ are connected by a
superedge if at least one node in $b$ is adjacent to a node in $b'$. From the relation between the
FPTs of $\G$ and $\GR$, the authors of Ref.~\cite{Gallos07} showed
\begin{equation}
T(l;k,k')=k^{\dw/\dk}f\left(\frac{l}{k^{1/\dk}}\right),
\label{eq:4}
\end{equation}
where $\dk$ is the exponent in the relation between the highest degree $K_{b}$ in box $b$ and the
supernode degree $k_{b}$ in $\GR$, i.e., $k_{b}\propto \lB^{-\dk}K_{b}$, and $f$ is a scaling
function. A similar expression to Eq.~(\ref{eq:4}) holds for $k'$. They have confirmed
numerically the validity of Eq.~(\ref{eq:4}) for several real-world networks and mathematical
models. From Fig.~2 of Ref.~\cite{Gallos07}, which displays the profile of the scaling function
$f(x)$ for confirming Eq.~(\ref{eq:4}), we notice that $f(x)$ is proportional to $x^{\dw}$. This
observation leads to an important fact for our work. Substituting $f(x)\propto x^{\dw}$ into
Eq.~(\ref{eq:4}), $T(l;k,k')$ is found to be actually independent of $k$ and $k'$ and $T(l)\propto
l^{\dw}$ for any $k$. In this case, $T(l)$ between two nodes separated by $l$ in $\G$ is
related to the FPT $\TR(l')$ between two supernodes separated by $l'=l/\lB$ in the renormalized
network $\GR$ by $T(l)=\TB\TR(l')$, where $\TB$ is the FPT from one end to the other of a covering
box. Due to the fractality of $\G$ and the lack of characteristic time in random walks, $\TR$ has
the same functional form as $T$. Thus, we have
\begin{equation}
T(l)=\TB T\left(\frac{l}{\lB}\right) .
\label{eq:5}
\end{equation}
The solution of this functional equation is $T(l)\propto l^{\dw}$, where the walk dimension $\dw$
is given by
\begin{equation}
\dw=\frac{\log \TB}{\log \lB} .
\label{eq:6}
\end{equation}
We should note that $\dw$ is independent of the degrees of the source and target nodes, which
implies that the bifractal nature of $\G$ does not influence the walk dimension.

The walk dimension $\dw$ for a fractal object embedded in Euclidean space also describes the mean
square displacement $\langle r^{2}(t)\rangle$ as $\langle r^{2}(t)\rangle\propto t^{2/\dw}$
\cite{ben-Avraham00}. Even in a network where the Euclidean distance is not defined, we have a
similar scaling for the mean topological displacement (MTD) defined by
$L_{i}(t)\equiv \sum_{j}p_{ij}(t)l_{ij}$, which corresponds to $\sqrt{\langle r^{2}(t)\rangle}$
\cite{Baronchelli08}. The fact that the bifractality of $\G$ is irrelevant to $\dw$ implies that
the MTD behaves as
\begin{equation}
L_{i}(t)\propto t^{1/\dw} ,
\label{eq:7}
\end{equation}
independently of the starting node $i$, where $\dw$ is given by Eq.~(\ref{eq:6}).

Next, we consider the spectral dimension $\ds$ of a bifractal network $\G$. The dimension $\ds$
describes the time dependence of the return probability as $p_{ii}(t)\propto t^{-\ds/2}$. The
return probability $p_{ii}(t)$ is inversely proportional to the number $V_{i}(t)$ of distinct nodes
visited by a walker starting from node $i$ during the first $t$ steps. By means of the MTD
$L_{i}(t)$, the visited volume $V_{i}(t)$ is given by $V_{i}(t)\propto [L_{i}(t)]^{\df}$, where
$\df$ is the fractal dimension near the node $i$ in $\G$. Thus, the proportionality
$p_{ii}(t)\propto [L_{i}(t)]^{-\df}$ and Eq.~(\ref{eq:7}) lead to the well-known relation
\cite{Rammal83}
\begin{equation}
\ds=\frac{2\df}{\dw} .
\label{eq:8}
\end{equation}
It should be emphasized that $\df$ in this relation represents the local fractal property in the
vicinity of the starting node $i$, but not the global fractal dimension $\Df$. Since the walk
dimension $\dw$ does not depend on $i$, the spectral dimension takes two values,
\begin{subequations}
\begin{align}
\dsmin&= \frac{2\dfmin}{\dw} , \label{eq:9a} \\[3pt]
\dsmax&= \frac{2\dfmax}{\dw} . \label{eq:9b}
\end{align}
\label{eq:9}
\end{subequations}
From the correspondence between the local structures and local fractal dimensions
\cite{Yamamoto23}, which is mentioned below Eq.~(\ref{eq:2}), we suppose that the return
probability of a random walker starting from an infinite-degree hub node is described by $\dsmin$,
while that of a random walker starting from a finite-degree non-hub node infinitely distant from
the hubs is characterized by $\dsmax$. The dimension $\dsmax$ also determines the mean return
probability $\langle p_{ii}(t)\rangle_{i}$, because the fraction of infinite-degree hub nodes is
infinitesimal in $\G$. Thus, $\dsmax$ always coincides with the global spectral dimension $\Ds$,
just as $\dfmax$ is equal to $\Df$. From the relation $p_{ii}(t)\propto 1/V_{i}(t)$, it is obvious
that the time dependence of the visited volume $V_{i}(t)$ is also described by two local spectral
dimensions. It is our main conclusion that $\dw$ remains constant regardless of the degree of the
starting node and that the spectral dimension $\ds$ takes two values as shown by Eq.~(\ref{eq:9})
due to the influence of multifractality of $\G$.

Let us interpret the walk and spectral dimensions of a bifractal network $\G$ in the context of
information diffusion. The fact that $\dw$ characterizing the MTD does not depend on the degree of
the starting node implies that the information spreading speed, in terms of how far the information
reaches within a given time, remains unchanged regardless of whether it originates from a
high-degree influencer or a low-degree non-influencer. On the other hand, the fact that $\dsmin$
determining the visited volume $V_{i}(t)$ is smaller than $\dsmax$ implies that the information
spreading speed, in the sense of how many people are informed within a given time, is faster when
it originates from the non-influencer rather than from the influencer.

\section{Walk and spectral dimensions of bifractal networks}
\label{sec:3}

Our general argument in the previous section predicts that the walk dimension $\dw$ of a bifractal
network takes a unique value irrespective of the starting node of a random walk, while the spectral
dimension takes two values, $\dsmin$ and $\dsmax$, depending on the starting node. In this section,
we analytically calculate the dimensions $\dw$, $\dsmin$, and $\dsmax$ for specific FSFNs formed by
two different types of models and numerically confirm these analytical predictions.

\subsection{Generator model}
\label{subsec:3-1}

A broad range of FSFNs can be synthetically constructed by using the generator model
\cite{Yamamoto23,Yakubo22}. In this model, we prepare a set of $s$ small graphs called
\textit{generators} $\{\Ggen^{(1)},\Ggen^{(2)},\dots,\Ggen^{(s)}\}$ in each of which two
non-adjacent nodes are designated as the \textit{root nodes}. For simplicity, these generators are
assumed to be symmetric with respect to the root nodes. This means that a subgraph of $\Ggen^{(i)}$
obtained by removing one root node from $\Ggen^{(i)}$ has the same topology as the subgraph formed
by removing another root node from $\Ggen^{(i)}$. We can easily extend the model to the case of
asymmetric generators \cite{Yakubo22}. The degree of the root node of at least one generator must
be $2$ or higher to ensure the scale-free property of the resulting network. Starting from a graph
$\G_{0}$ with two nodes and a single edge connecting them, the $\gen$-th generation network
$\G_{\gen}$ is recursively constructed by replacing every edge in $\G_{\gen-1}$ with one of the
multiple generators $\Ggen^{(i)}$ with the predefined probability $p^{(i)}$, where
$\sum_{i=1}^{s}p^{(i)}=1$. In the edge replacement procedure, two root nodes in the selected
generator coincide with the two end nodes of the replaced edge.

\begin{figure}[tttt]
    \begin{center}
    \includegraphics[width=0.98\linewidth]{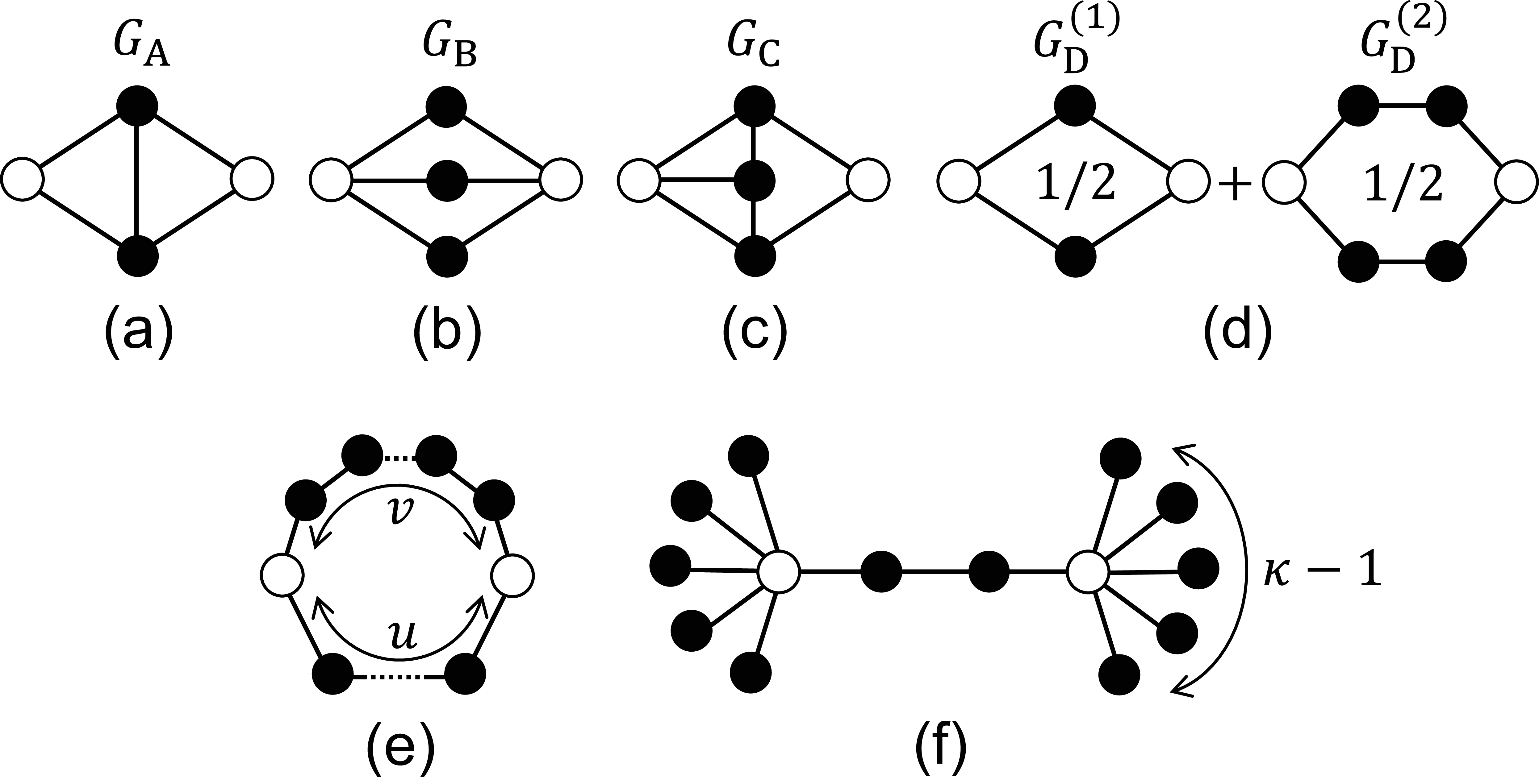}
    \caption{
Examples of generators. Two white nodes in each graph represent the root nodes. Graphs in the top
row show (a) symmetric single generator $G_{\text{A}}$, (b) symmetric single generator
$G_{\text{B}}$, (c) asymmetric single generator $G_{\text{C}}$, and (d) symmetric two generators
$G_{\text{D}}^{(1)}$ and $G_{\text{D}}^{(2)}$ to construct FSFNs used for numerical confirmation.
The fraction ``$1/2$" indicated inside the graph is the edge replacement probability. Two graphs in
the bottom row give the generators (e) for the $(u,v)$-flower \cite{Rozenfeld07} and (f) for the
Song-Havlin-Makse model \cite{Song06}.
    }
    \label{fig:1}
    \end{center}
\end{figure}

A network $\G$ formed by the above generator model after a sufficient number of edge replacements
possesses scale-free and fractal properties. The degree exponent $\gamma$ defining the
power-law behavior of the degree distribution $P(k)$ is given by Ref.~\cite{Yamamoto23}
\begin{equation}
\gamma=1+\frac{\log \mgenav}{\log \kappaav} .
\label{eq:10}
\end{equation}
Here, $\mgenav=\sum_{i=1}^{s}\mgeni p^{(i)}$ and $\kappaav=\sum_{i=1}^{s}\kappa^{(i)}p^{(i)}$ are
the mean number of edges and the mean degree of a root node of the multiple generators, where
$\mgeni$ and $\kappa^{(i)}$ denote the number of edges and the degree of the root node of
$\Ggen^{(i)}$. On the other hand, the fractal dimension of $\G$ is presented by
\begin{equation}
\Df=\frac{\log \mgenav}{\log \lambdaeff} ,
\label{eq:11}
\end{equation}
where $\lambdaeff$ is the effective inter-root-node distance, which is not the simple average
$\sum_{i=1}^{s}\lambda^{(i)}p^{(i)}$ of the distance $\lambda^{(i)}$ between the root nodes of each
generator $\Ggen^{(i)}$. A method to calculate $\lambdaeff$ from the set $\{\lambda^{(i)}\}$ is
presented in Ref.~\cite{Yamamoto23}. We should emphasize that in the case of a single generator
$\Ggen$ (namely, $s=1$ and $p^{(1)}=1$), $\lambdaeff$ is just equal to the inter-root-node distance
$\lambda$ of $\Ggen$. Furthermore, FSFNs formed by the generator model satisfy Eq.~(\ref{eq:1}),
and thus, possess bifractal structures \cite{Yamamoto23}. The local fractal dimensions are given by
\begin{subequations}
\begin{align}
\dfmin&= \frac{\log (\mgenav/\kappaav)}{\log \lambdaeff} , \label{eq:12a}\\
\dfmax&= \frac{\log \mgenav}{\log \lambdaeff} . \label{eq:12b}
\end{align}
\label{eq:12}
\end{subequations}
If we employ the symmetric single generator shown by Fig.~\ref{fig:1}(e), where the root nodes
partition the cycle of length $(u+v)$, we can reproduce the $(u,v)$-flowers \cite{Rozenfeld07}. The
generator depicted in Fig.~\ref{fig:1}(f), where the root nodes of degree $\kappa$ are connected to
each other via a single path of length $3$, gives FSFNs formed by the Song-Havlin-Makse model
\cite{Song06}. These examples demonstrate that the generator model is a generalization of various
existing models of FSFNs.

Let us calculate the walk dimension $\dw$ of an $\gen$-th generation FSFN, $\G_{\gen}$, formed by
the generator model. To this end, we consider the FPT between the renormalized root nodes (RRNs) in
$\G_{\gen}$, where the RRNs are the oldest two nodes that constituted $\G_{0}$. We utilize the
general relation $T_{ij}+T_{ji}=2MR_{ij}$ for a network with $M$ edges \cite{Chandra96}, where
$T_{ij}$ is the FPT from node $i$ to $j$ and $R_{ij}$ is the resistance between $i$ and $j$ when
each edge is a conductor of unit resistance. For $\G_{\gen}$ formed by a single symmetric
generator, the FPT from one RRN to the other is obviously the same as the FPT in the opposite
direction. Thus, the FPT $T_{\gen}$ between the RRNs in $\G_{\gen}$ with $M_{\gen}$ edges is given
by
\begin{equation}
T_{\gen}=M_{\gen} R_{\gen} ,
\label{eq:13}
\end{equation}
where $R_{\gen}$ is the resistance between the RRNs in $\G_{\gen}$. Even in the case of multiple
generators and/or when some of them are asymmetric, $T_{\gen}$ from one RRN to the other is
statistically the same as that in the opposite direction for sufficiently large $\gen$, because the
choice of generators and their orientations (if asymmetric) at the edge replacement is random.
Hence, Eq.~(\ref{eq:13}) holds for any type of generator.

\begin{table*}
\caption{\label{tab:1}
Walk dimension $\dw$, spectral dimensions $\dsmin$ and $\dsmax$, and other related quantities for
FSFNs formed by the generators shown in Figs.~\ref{fig:1}(a)--\ref{fig:1}(d). The definitions of
$\Ds$, $\mgenav$, $\kappaav$, $\lambdaeff$, $\Rrooteff$, $\gamma$, $\dfmin$, $\dfmax$, and $\Df$
are presented in the main text.}
 \begin{ruledtabular}
  \begin{tabular}{lcccccccccc}
  Generator(s)
    & $\dw$    & $\dsmin$ & $\dsmax(=\Ds)$  & $\mgenav$ & $\kappaav$ & $\lambdaeff$ & $\Rrooteff$ & $\gamma$ & $\dfmin$ & $\dfmax(=\Df)$  \\[2pt] \hline
  $G_{\text{A}}$
    & $2.3219$ & $1.1386$ & $2$             & $5$       & $2$        & $2$          & $1$         & $3.3219$ & $1.3219$ & $2.3219$  \rule[0mm]{0mm}{4mm} \\
  $G_{\text{B}}$
    & $2$      & $1$      & $2.5849$        & $6$       & $3$        & $2$          & $2/3$       & $2.6309$ & $1$      & $2.5849$  \\
  $G_{\text{C}}$
    & $2.6147$ & $1.1362$ & $2.1474$        & $7$       & $5/2$      & $2$          & $7/8$       & $3.1237$ & $1.4854$ & $2.8074$  \\
  $G_{\text{D}}^{(1)}+G_{\text{D}}^{(2)}$
    & $2.1616$ & $1.0045$ & $1.7644$        & $5$       & $2$        & $2.3256$     & $1.2398$    & $3.3219$ & $1.0857$ & $1.9069$
  \end{tabular}
 \end{ruledtabular}
\end{table*}
The FPT for $\G_{\gen}$ formed by a single symmetric generator $\Ggen$ is examined at first (i.e.,
$s=1$ and $p^{(1)}=1$). In this case, $M_{\gen}$ and $R_{\gen}$ are related to $M_{\gen-1}$ and
$R_{\gen-1}$ of $\G_{\gen-1}$ by the relations
\begin{gather}
M_{\gen}=\mgen M_{\gen-1} , \label{eq:14}\\
R_{\gen}=\Rroot R_{\gen-1} , \label{eq:15}
\end{gather}
where $\mgen$ is the number of edges in $\Ggen$ and $\Rroot$ is the resistance between the root
nodes in $\Ggen$. Therefore, we have
\begin{equation}
T_{\gen}=\mgen\Rroot T_{\gen-1} .
\label{eq:16}
\end{equation}
On the other hand, the shortest-path distance $l_{\gen}$ between the RRNs in $\G_{\gen}$ is related
to $l_{\gen-1}$ for $\G_{\gen-1}$ by
\begin{equation}
l_{\gen}=\lambda l_{\gen-1} ,
\label{eq:17}
\end{equation}
where $\lambda$ is the distance between the root nodes of $\Ggen$. Equations (\ref{eq:16}) and
(\ref{eq:17}) provide the relation,
\begin{equation}
T(\lambda l)=\mgen\Rroot T(l) ,
\label{eq:18}
\end{equation}
where $T(l)$ is the FPT between the RRNs separated by $l$ from each other. The solution of the
above functional equation is $T(l)\propto l^{\dw}$ with
\begin{equation}
\dw=\frac{\log \left(\mgen\Rroot \right)}{\log \lambda} .
\label{eq:19}
\end{equation}
We should note that the quantity $\mgen\Rroot$ in the numerator means the FPT between the root
nodes in $\Ggen$.

Even for the generator model with multiple generators, $\dw$ can be calculated in basically the
same way. The recurrence equations for $M_{\gen}$, $R_{\gen}$, and $l_{\gen}$ are, however,
slightly different from Eqs.~(\ref{eq:14}), (\ref{eq:15}), and (\ref{eq:17}) for the
single-generator model. If $\G_{\gen-1}$ contains $M_{\gen-1}$ edges, $p^{(i)}M_{\gen-1}$ edges in
$\G_{\gen-1}$ are replaced with the $i$-th generator $\Ggen^{(i)}$ and proliferate to
$p^{(i)}M_{\gen-1}m_{\text{gen}}^{(i)}$ edges in $\G_{\gen}$. Thus, the total number of edges in
$\G_{\gen}$ is given by $M_{\gen-1}\sum_{i}^{s}m_{\text{gen}}^{(i)}p^{(i)}$, and we have
\begin{equation}
M_{\gen}=\mgenav M_{\gen-1} ,
\label{eq:20}
\end{equation}
instead of Eq.~(\ref{eq:14}). Similarly, for the multigenerator model, the resistance $R_{\gen}$
and the distance $l_{\gen}$ between the RRNs are, for $n\gg 1$, presented by
\begin{align}
R_{\gen}&=\Rrooteff R_{\gen-1} ,
\label{eq:21} \\
l_{\gen}&=\lambdaeff l_{\gen-1} ,
\label{eq:22}
\end{align}
instead of Eqs.~(\ref{eq:15}) and (\ref{eq:17}), respectively. Here, $\Rrooteff$ is the effective
resistance between root nodes in multiple generators. As in the case of $\lambdaeff$, $\Rrooteff$
is not a simple average of the inter-root-node resistance $\Rroot^{(i)}$ of $\Ggen^{(i)}$, because
edge replacements change the series, parallel, and bridging configurations in composite resistance.
The actual calculation method of $\Rrooteff$ is explained in the Appendix. From Eqs.~(\ref{eq:20}),
(\ref{eq:21}), and (\ref{eq:22}), we have the relation $T(\lambdaeff l)=\mgenav\Rrooteff T(l)$ as
the multigenerator version of Eq.~(\ref{eq:18}), and finally obtain
\begin{equation}
\dw=\frac{\log \left(\mgenav\Rrooteff \right)}{\log \lambdaeff} .
\label{eq:23}
\end{equation}
Since $\mgenav$, $\lambdaeff$, and $\Rrooteff$ are equal to $\mgen$, $\lambda$, and $\Rroot$,
respectively, for $s=1$, $\dw$ given by Eq.~(\ref{eq:23}) includes that by Eq.~(\ref{eq:19}) as a
special case.

The spectral dimensions $\dsmin$ and $\dsmax$ for the generator model immediately follow from the
above $\dw$. Substituting Eqs.~(\ref{eq:12}) and (\ref{eq:23}) into Eq.~(\ref{eq:9}), these
dimensions are given by
\begin{subequations}
\begin{align}
\dsmin&= \frac{2\log \left(\mgenav/\kappaav\right)}{\log \left(\mgenav \Rrooteff\right)} , \label{eq:24a}\\
\dsmax&= \frac{2\log \mgenav}{\log \left(\mgenav \Rrooteff\right)} . \label{eq:24b}
\end{align}
\label{eq:24}
\end{subequations}
Given the structural information of generators, we can calculate $\dw$, $\dsmin$, and $\dsmax$
analytically by using Eqs.~(\ref{eq:23}) and (\ref{eq:24}). Actual values of these dimensions and
related quantities for FSFNs formed by the generators shown in Figs.~\ref{fig:1}(a)--\ref{fig:1}(d)
are summarized in Table~\ref{tab:1}.

\begin{figure}
\begin{center}
\includegraphics[width=0.44\textwidth]{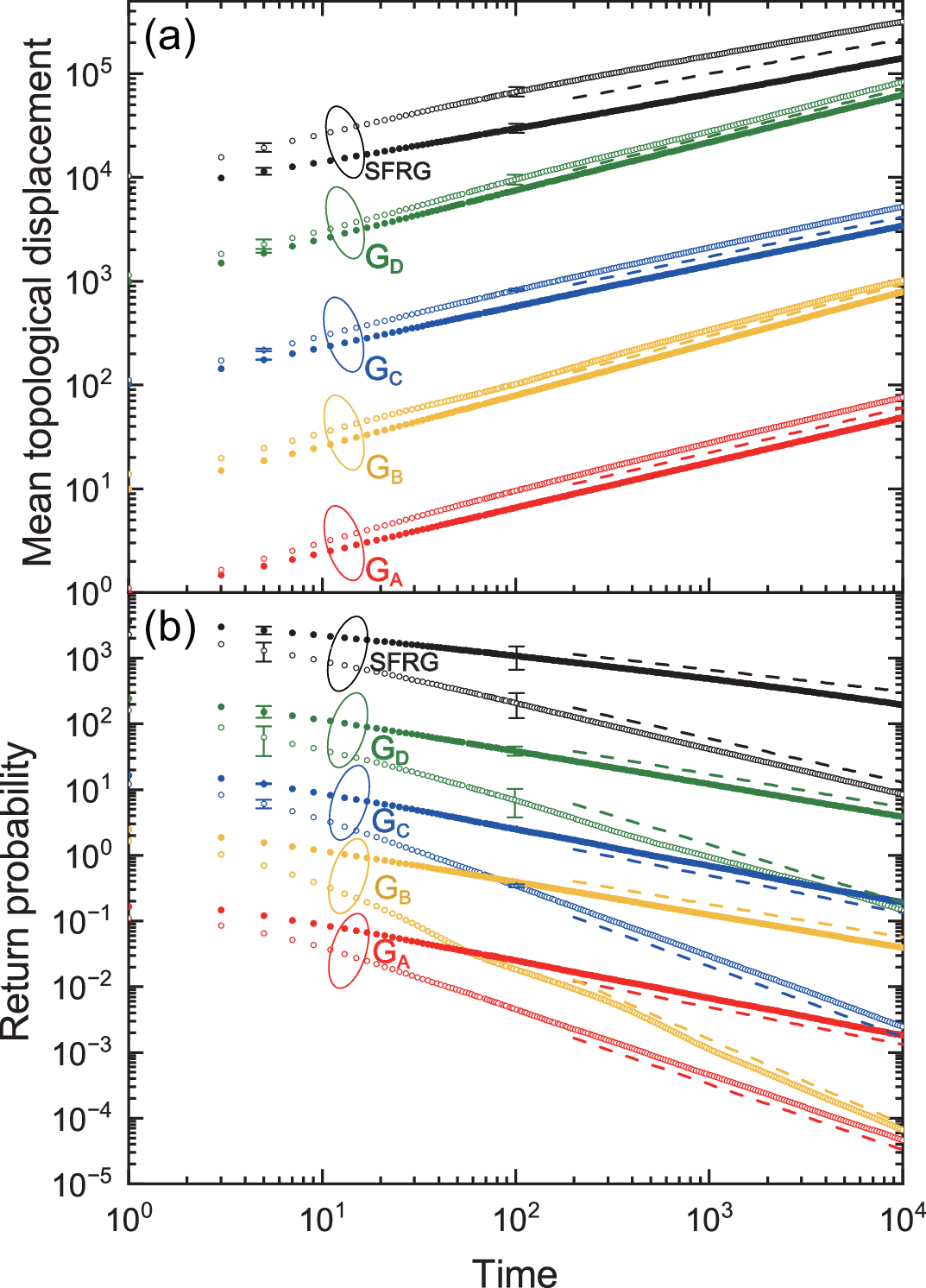}
\caption{
Numerical simulations on (a) the MTD $L_{i}(t)$ and (b) the return probability $p_{ii}(t)$. In each
panel, four (red, yellow, blue, and green) pairs of filled and open symbols from the bottom are the
results for FSFNs formed by the generator model with the generators labeled near the symbols. The
top (black) pair gives the results for scale-free random graphs at the percolation critical point.
Filled and open symbols represent the results starting from hub and non-hub nodes, respectively.
Typical magnitudes of error bars are shown at $t=5$ and $100$. Dashed lines close to symbols
indicate the slopes by corresponding theoretical predictions. For graphical reasons, the results of
the black (SFRG), green ($G_{\text{D}}$), blue ($G_{\text{C}}$), and yellow ($G_{\text{B}}$) pairs
are shifted upward by $10^{4}$, $10^{3}$, $10^{2}$, and $10$, respectively.
}
\label{fig:2}
\end{center}
\end{figure}

Let us confirm the above analytical predictions by numerically solving the master equation
Eq.~(\ref{eq:3}). Figure 2 shows the time dependences of the MTD $L_{i}(t)$ [Fig.~\ref{fig:2}(a)]
and the return probability $p_{ii}(t)$ [Fig.~\ref{fig:2}(b)]. Filled and open symbols represent the
numerical results for random walks starting from hub and non-hub nodes, respectively. The starting
hub node in each network is specified as a node selected randomly from the highest degree
($k_{\text{max}}$) nodes, while the starting non-hub node is a node selected at random from the
lowest degree ($k_{\text{min}}$) nodes far away from high degree nodes. In each panel of
Fig.~\ref{fig:2}, four pairs of filled and open symbols from the bottom show the results for FSFNs
formed by the generators $G_{\text{A}}$, $G_{\text{B}}$, $G_{\text{C}}$, and
$G_{\text{D}}^{(1)}+G_{\text{D}}^{(2)}$ presented in Figs.~\ref{fig:1}(a)--\ref{fig:1}(d). Although
FSFNs by the symmetric single generators $G_{\text{A}}$ and $G_{\text{B}}$ are deterministic, those
by the asymmetric generator $G_{\text{C}}$ and multigenerator
$G_{\text{D}}^{(1)}+G_{\text{D}}^{(2)}$ have stochastic structures. We then take the average over
$100$ realizations of such stochastic FSFNs. Fundamental characteristics of the constructed FSFNs
are summarized in Table \ref{tab:2}.
\begin{table}[b!]
\caption{\label{tab:2}
Characteristics of FSFNs used for numerical calculations, that are formed by the generator model
with the generators $G_{\text{A}}$, $G_{\text{B}}$, $G_{\text{C}}$, and
$G_{\text{D}}^{(1)}+G_{\text{D}}^{(2)}$ presented in Figs.~\ref{fig:1}(a)--\ref{fig:1}(d). The
symbols $\gen$, $N$, $M$, $k_{\text{min}}$, and $k_{\text{max}}$ denote the generation, number of
nodes, number of edges, and minimum and maximum degrees, respectively. Noninteger values imply the
averages over network samples.
}
 \begin{ruledtabular}
  \begin{tabular}{lccccc}
  Generator       & $\gen$ & $N$           & $M$           & $k_{\text{min}}$ & $k_{\text{max}}$  \\[2pt] \hline
  $G_{\text{A}}$
                  & $8$    & $195\,314$    & $390\,625$    & $3$              & $384$             \rule[0mm]{0mm}{4mm} \\
  $G_{\text{B}}$
                  & $8$    & $1\,007\,771$ & $1\,679\,616$ & $2$              & $6\,561$          \\
  $G_{\text{C}}$
                  & $8$    & $2\,882\,402$ & $5\,764\,801$ & $3$            & $2\,159.25$       \\
  $G_{\text{D}}^{(1)}+G_{\text{D}}^{(2)}$
                  & $8$    & $288\,567.42$ & $384\,764.16$ & $2$              & $256$
  \end{tabular}
 \end{ruledtabular}
\end{table}
The time dependence of the MTD gives the walk dimension $\dw$ as shown by Eq.~(\ref{eq:7}). The
results of red, yellow, blue, and green pairs of symbols in Fig.~\ref{fig:2}(a) clearly show
that $\dw$ does not depend on whether the starting node is hub or non-hub. Dashed straight lines
between filled and open symbols reflect the theoretical values of $\dw$ for the corresponding
networks, which are presented in Table \ref{tab:1}. The slopes of our numerical results agree quite
well with those of dashed lines. In contrast, the return probability $p_{ii}(t)$ shown in
Fig.~\ref{fig:2}(b) strongly depends on the starting node $i$, which implies that the spectral
dimension $\ds$ takes different values in the vicinity of the hub and non-hub nodes, because of the
relation $p_{ii}(t)\propto t^{-\ds/2}$. In addition, our numerical results coincide with the
theoretical predictions shown by dashed lines. All these numerical results support the validity of
the analytical argument presented in Sec.~\ref{sec:2}.

\subsection{Scale-free random graphs at criticality}
\label{subsec:3-2}

The generator model presented above constructs FSFNs in a synthetic manner, and the resulting
networks exhibit clear hierarchical structures. In this subsection, as an example of nonsynthetic
FSFNs, we consider the giant connected components of uncorrelated scale-free networks at the
percolation critical point and examine their walk and spectral dimensions. When the degree
distribution of a scale-free random graph (SFRG) is proportional to $k^{-(\gamma+1)}$ for high
degree $k$, its giant component (GC) at criticality has a power-law degree distribution with the
degree exponent $\gamma$ \cite{Bialas08} and exhibits a fractal structure with the fractal
dimension $\Df=(\gamma-1)/(\gamma-2)$ for $2< \gamma < 3$ and $\Df=2$ for $\gamma\ge 3$
\cite{Cohen03}. Thus, the GC of an SFRG at criticality is an FSFN without explicit hierarchical
nesting. Hereafter, we refer to the GC of an SFRG at the percolation critical point as ``critical
SFRG" shortly. Recently, it has been elucidated that critical SFRGs possess the bifractal nature in
their structures \cite{Yamamoto23}. The local fractal dimensions are given by
\begin{equation}
\dfmin=
\begin{cases}
1                                            & \text{for } 2<\gamma<3 \ ,\\[8pt]
\displaystyle \frac{2(\gamma-2)}{\gamma-1}   & \text{for } \gamma\ge 3 \ ,
\end{cases}
\label{eq:25}
\end{equation}
and
\begin{equation}
\dfmax=
\begin{cases}
\displaystyle \frac{\gamma-1}{\gamma-2} & \text{for } 2<\gamma<3 \ ,\\[8pt]
2                                       & \text{for } \gamma\ge 3 \ .
\end{cases}
\label{eq:26}
\end{equation}
Thus, we expect two values of spectral dimensions $\dsmin$ and $\dsmax$ for a critical SFRG.

The (global) spectral dimension $\Ds$ of a percolation network at criticality has been studied more
than 40 years ago by Alexander and Orbach \cite{Alexander82}. They showed that $\Ds$ is equal to
$4/3$ for a critical percolation network on an infinite dimensional lattice. Since the percolation
transition of an SFRG with $\gamma+1\ge 4$ is known to belong to the same universality class as
that on infinite dimensional lattice \cite{Cohen02}, we can consider that $\Ds$ of critical SFRGs
with $\gamma\ge 3$ is equal to $4/3$. On the other hand, a field theoretical approach to the
network analysis has revealed that $\Ds$ of a scale-free random branching tree at the critical
point is given by $\Ds=2(\gamma-1)/(2\gamma-3)$ for $2<\gamma<3$, where $\gamma$ is the degree
exponent of the critical branching tree \cite{Burda01}. Since the structure of the critical
scale-free random branching tree with the degree exponent $\gamma$ is equivalent to that of a
critical SFRG with the exponent $\gamma$, we can conclude that $\Ds$ of the critical SFRG with
$\gamma$ less than $3$ is given by
\begin{equation}
\Ds=\frac{2(\gamma-1)}{2\gamma-3} \quad \text{for $2<\gamma<3$}.
\label{eq:27}
\end{equation}
It should be emphasized that the above spectral dimensions are nothing but the global dimensions.
Recalling that the local spectral dimension $\dsmax$ is always equal to the global dimension, we
obtain
\begin{equation}
\dsmax=
\begin{cases}
\displaystyle \frac{2(\gamma-1)}{2\gamma-3} & \text{for } 2<\gamma<3 \ ,\\[8pt]
\displaystyle \frac{4}{3}                   & \text{for } \gamma\ge 3 \ .
\end{cases}
\label{eq:28}
\end{equation}
The walk dimension $\dw$ can be calculated by substituting this result and $\dfmax$ given by
Eq.~(\ref{eq:26}) into Eq.~(\ref{eq:9b}), then we have
\begin{equation}
\dw=
\begin{cases}
\displaystyle \frac{2\gamma-3}{\gamma-2} & \text{for } 2<\gamma<3 \ ,\\[8pt]
\displaystyle 3                          & \text{for } \gamma\ge 3 \ ,
\end{cases}
\label{eq:29}
\end{equation}
regardless of the starting node of a random walk. We then calculate the spectral dimension $\dsmin$
by substituting this dimension $\dw$ and $\dfmin$ given by Eq.~(\ref{eq:25}) into Eq.~(\ref{eq:9a})
as
\begin{equation}
\dsmin=
\begin{cases}
\displaystyle \frac{2(\gamma-2)}{2\gamma-3}    & \text{for } 2<\gamma<3 \ ,\\[8pt]
\displaystyle \frac{4(\gamma-2)}{3(\gamma-1)}  & \text{for } \gamma\ge 3 \ .
\end{cases}
\label{eq:30}
\end{equation}
The dimension $\dsmin$ becomes $4/3$ in the limit of $\gamma\to \infty$, which is the same as
$\dsmax$. This is due to the loss of bifractality in non-scale-free networks.

In order to verify the above theoretical prediction, SFRGs at criticality are numerically
constructed by the configuration model \cite{Newman01,Bender78}. The degree distribution of SFRGs
is chosen as
\begin{equation}
P(k)=
\begin{cases}
\displaystyle \frac{c}{k^{\gamma+1}+d^{\gamma+1}} & \text{for } 1\le k\le 10^{3} \ ,\\[8pt]
0 & \text{otherwise} \ ,
\end{cases}
\label{eq:31}
\end{equation}
where $\gamma$ is fixed at $\gamma=3$, $d$ is a parameter, and $c$ is the normalization constant
depending on $d$. This distribution function asymptotically follows $P(k)\propto k^{-(\gamma+1)}$
for $k\gg d$. The parameter $d$ alters the profile of $P(k)$ mainly at $k\lesssim d$, so by varying
the value of $d$, we can control the moments of $P(k)$ without changing the degree exponent. If the
ratio $\langle k^{2}\rangle/\langle k\rangle$ obtained by this distribution is equal to $2$, SFRGs
have critical structures \cite{{Molloy95}}. In practice, this ratio is chosen as $\langle
k^{2}\rangle/\langle k\rangle=2.05$, which is realized by $d=1.3766$, to prevent GC sizes from
becoming too small. We prepare $500$ realizations of SFRGs with $N=10^{6}$ nodes and extract the
largest connected components as GCs. The average, minimum, and maximum GC sizes are $18\,015.2$,
$10\,005$, and $36\,879$, respectively. The starting hub and non-hub nodes of random walks on each
GC are selected in the same way as the case of the generator model. The numerical results of the
MTD and the return probability averaged over these GCs are presented at the top of
Figs.~\ref{fig:2}(a) and \ref{fig:2}(b), respectively. The theoretical values of the walk and
spectral dimensions for $\gamma=3$ are $\dw=3$, $\dsmin=2/3$, and $\dsmax=4/3$, and the slopes
corresponding to these values are indicated by dashed lines. Our numerical results clearly
demonstrate the validity of the theoretical predictions Eqs.~(\ref{eq:28})--(\ref{eq:30}).

\section{Discussion and Conclusion}
\label{sec:4}

The fact that the spectral dimension near a hub node in a scale-free network takes a
different value from the global dimension $\Ds$ has already been pointed out by previous work
\cite{Hwang12}. The authors of Ref.~\cite{Hwang12} explored a mean-field solution of
Eq.~(\ref{eq:3}) based on a heuristic argument and calculated the time dependence of the return
probability. They obtained the local spectral dimension $\dshub$ near a hub node as
\begin{equation}
\dshub=\Ds \frac{\gamma-2}{\gamma-1} .
\label{eq:32}
\end{equation}
This dimension coincides with our $\dsmin$ given by Eq.~(\ref{eq:9a}), which can be easily
confirmed by substituting $\dw=2\dfmax/\dsmax$ from Eq.~(\ref{eq:9b}) into Eq.~(\ref{eq:9a}) and
using Eq.~(\ref{eq:2a}), $\dfmax=\Df$, and $\dsmax=\Ds$. Their argument does not explicitly assume
that networks are fractal, but rather that they are \textit{uncorrelated}. Nevertheless, most of
the numerical confirmations in Ref.~\cite{Hwang12} have been conducted for \textit{correlated} FSFNs.
Thus, it is not clear what kind of networks are actually characterized by $\dshub$. Furthermore, it
also remains to be elucidated what structural properties in the vicinity of hubs are related to the
fact that $\dshub\ne \Ds$. Our present work provides a sufficient condition for $\dsmin\ne \dsmax$
and clarifies the structural origin of this phenomenon. Namely, we found that $\dsmin$ takes a
different value from $\dsmax(=\Ds)$ at least in bifractal networks, and this is due to the
difference in local fractality around the hub and non-hub nodes.

The same authors as in Ref.~\cite{Hwang12} have demonstrated without assuming the uncorrelatedness
of networks that the spectral dimension near the highest degree nodes in $(u,v)$-flowers differs
from the global dimension, and calculated analytically the local spectral dimension around the hubs
by using a generating function-based renormalization-group approach \cite{Hwang13}. As mentioned in
Sec.~\ref{subsec:3-1}, the $(u,v)$-flower model is a special case of the single-generator model,
and a flower is constructed by employing the generator shown in Fig.~\ref{fig:1}(e). Thus, the
spectral dimension $\dsmin$ is calculated by substituting $\mgenav=u+v$, $\kappaav=2$, and
$\Rrooteff=uv/(u+v)$ into Eq.~(\ref{eq:24a}), and we have
\begin{equation}
\dsmin= \frac{2\log \left[(u+v)/2\right]}{\log \left(uv\right)} .
\label{eq:33}
\end{equation}
This expression is exactly the same as the result in Ref.~\cite{Hwang13}. Since Eq.~(\ref{eq:24})
is a direct consequence of the bifractality of FSFNs, our derivation is much more straightforward,
and enables us to easily compute $\dsmin$ and $\dsmax$ for a wide range of synthetic hierarchical
FSFNs, including $(u,v)$-flowers.

\begin{figure}
\begin{center}
\includegraphics[width=0.44\textwidth]{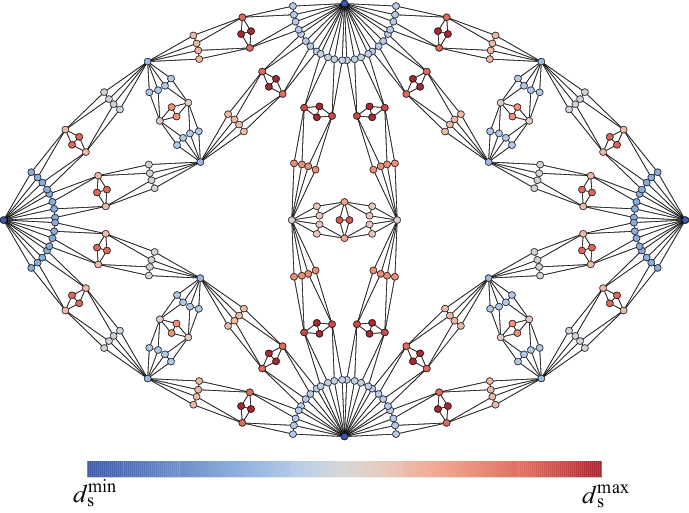}
\caption{
Local spectral dimension around each node in an FSFN. The value of $\ds$ is indicated by color
[from blue for $\ds=\dsmin(=1.1386)$ to red for $\ds=\dsmax(=2)$]. The network is formed by
the symmetric single-generator model with the generator $G_{\text{A}}$ shown in Fig.~\ref{fig:1}(a)
in the 4th generation. The numbers of nodes and edges are $N=314$ and $M=625$, respectively.
}
\label{fig:3}
\end{center}
\end{figure}
Finally, we confirm the relation between the local spectral dimensions, node degree, and node
position. Figure \ref{fig:3} represents the local spectral dimension $\ds$ around each node in the
4th generation network $\G_{4}$ formed by the generator model with the generator $G_{\text{A}}$
shown in Fig.~\ref{fig:1}(a). The value of $\ds$ on each node $i$ is obtained from the return
probability $p_{ii}(t)$, which is calculated by solving Eq.~(\ref{eq:3}) numerically. As expected,
the local spectral dimension $\ds$ on the highest degree nodes ($2$ nodes with $k=24$ located at
the top and bottom) and the second highest degree nodes ($2$ nodes with $k=16$ located on the left
and right sides) is close to $\dsmin(=1.1386)$, whereas $\ds$ is close to $\dsmax(=2)$ for the
lowest degree nodes far from hubs. Due to the finite-size effect of $\G_{4}$, $\ds$ cannot be
perfectly dichotomized into $\dsmin$ and $\dsmax$. There are many nodes with colors corresponding
to intermediate values of $\ds$. This does not imply, however, that local spectral dimensions on
these nodes actually take intermediate values between $\dsmin$ and $\dsmax$, but rather indicates
that the return probability $p_{ii}(t)$ exhibits a crossover behavior between $t^{-\dsmin/2}$ and
$t^{-\dsmax/2}$. A relevant crossover behavior in the return probability $p_{k}(t)$ has been
discussed by Ref.~\cite{Hwang12}, where $p_{k}(t)$ is the average of $p_{ii}(t)$ over starting
nodes with degree $k$. This previous work reveals that the crossover time $t_{\text{c}}$ of
$p_{k}(t)$ depends on the degree of the starting node as $t_{\text{c}}\sim k^{2(\gamma-1)/\Ds}$.
Our result shown in Fig.~\ref{fig:3} demonstrates that individual $p_{ii}(t)$'s depend not only on
$k$ but also on the relative position of the starting node in the network. Although we focused on
the asymptotic behavior of random walks in FSFNs, it is intriguing to elucidate how the structural
crossover in a bifractal network relates to the dynamical crossover in $p_{ii}(t)$, and
furthermore, how the crossover in $p_{ii}(t)$ shifts to that in $p_{k}(t)$ through the averaging
procedure.

In conclusion, we have studied random walks on FSFNs to understand how their structural
bifractality affects the dynamical properties of networks. We focused, in particular, on the walk
dimension $\dw$ and the spectral dimension $\ds$, which are fundamental dynamical dimensions
characterizing random walks. From the relation between the FPT in a fractal network $\G$ and that
in its renormalized network $\GR$ obtained by covering $\G$ with boxes of diameter $\lB$, it has
been found that the walk dimension is determined by the FPT of the covering box and the box
diameter independently of the starting node of a random walk. In contrast, the spectral dimension
takes two different values, $\dsmin$ and $\dsmax(> \dsmin)$, depending on the starting node, which
is a consequence of the bifractal nature of the FSFN. In the thermodynamic limit, $\dsmin$
describes random walks starting from infinite-degree hub nodes, while $\dsmax$ characterizes those
starting from finite-degree non-hub nodes infinitely distant from hub nodes. The dimension $\dsmax$
is equal to the global spectral dimension $\Ds$, i.e., $\ds$ averaged over all starting nodes. In
addition, we have provided analytical expressions for $\dw$, $\dsmin$, and $\dsmax$ for FSFNs
formed by the generator model and the giant components of critical scale-free random graphs. The
numerical verification of these expressions supports the validity of our analytical arguments.

It is imperative to understand the impact of structural bifractality on various dynamics in complex
networks, such as synchronization, epidemic spreading, opinion formation, and strategic games. A
possible approach is to study the spectral properties of the Laplacian matrix describing an FSFN,
because the spectral density and eigenvectors of the Laplacian matrix play crucial roles in these
dynamics and are closely related to the spectral dimension. It is also intriguing to investigate
whether the conclusions obtained in this paper hold for GCs at nontrivial percolation transitions,
such as critical structures in simplicial complexes \cite{Sun19}, hybrid transitions in planar
hyperbolic manifolds \cite{Kryven19}, and critical phases observed in hierarchical networks
\cite{Nogawa09,Hasegawa10,Boettcher12}. We believe that the present work can provide valuable
insights into such further investigations.

\begin{acknowledgements}
The authors thank S. Mizutaka for fruitful discussions.
This work was supported by JSPS KAKENHI Grant Number 22K03463.
\end{acknowledgements}

\appendix*
\section{Calculation of $\Rrooteff$}
\label{appendix:a}

The resistance $R_{\gen}$ between the RRNs of an FSFN $\G_{\gen}$ formed by the generator model
satisfies the relation $R_{\gen}=\Rrooteff R_{\gen-1}$, where $\Rrooteff$ is the effective
inter-root-node resistance of the generators. The coefficient $\Rrooteff$ is expected to become a
constant independent of $n$ for sufficiently large $n$. Here, we calculate $\Rrooteff$ for $n\gg 1$
from a given set of generators $\{ \Ggen^{(i)}|1\le i\le s\}$ whose edges are made of conductors
with unit resistance. According to the idea of a renormalization scheme, the $\gen$-th generation
network $\G_{\gen}$ takes a structure in which each edge of one of the generators is replaced with
one of the possible structures of $(\gen-1)$-th generation networks $\G_{\gen-1}$. Therefore, we
have
\begin{equation}
R_{\gen}=\sum_{i=1}^{s}p^{(i)}R_{\gen}^{(i)} ,
\label{eq:a1}
\end{equation}
where
\begin{equation}
R_{\gen}^{(i)}=F^{(i)}\left(\{R_{\gen-1}^{(j)}\},\{p^{(j)}\} \right) .
\label{eq:a2}
\end{equation}
In the above expression, $R_{\gen}^{(i)}$ is the average inter-root-node resistance of the
generator $\Ggen^{(i)}$ in which each edge has one of the resistances
$\{R_{\gen-1}^{(j)}|1\le j\le s\}$ with probability $p^{(j)}$, and $F^{(i)}$ is the function
mapping from the set $\{R_{\gen-1}^{(j)}\}$ to $R_{\gen}^{(i)}$. Given the structures of the
generators, the functions $\{F^{(i)}\}$ are uniquely determined, and we can compute
$R_{\gen}^{(i)}$ iteratively using Eq.~(\ref{eq:a2}). The resistance $R_{\gen}$ is then obtained by
Eq.~(\ref{eq:a1}). Since the relation $R_{\gen}=\Rrooteff R_{\gen-1}$ holds for $\gen\gg 1$, the
effective inter-root-node resistance $\Rrooteff$ is given by
\begin{equation}
\Rrooteff=\lim_{\gen\to\infty}\frac{R_{\gen}}{R_{\gen-1}} .
\label{eq:a3}
\end{equation}
Convergence of the right-hand side of Eq.~(\ref{eq:a3}) is usually very fast.

As an example, we calculate $\Rrooteff$ for the combination of two generators $G_{\text{D}}^{(1)}$
and $G_{\text{D}}^{(2)}$ shown in Fig.~\ref{fig:1}(d). Although the edge replacement probabilities
$p^{(1)}$ and $p^{(2)}$ for $G_{\text{D}}^{(1)}$ and $G_{\text{D}}^{(2)}$ are both $1/2$, we denote
these probabilities by $p$ and $1-p$, respectively, to clarify the meanings of expressions. In this
case, Eq.~(\ref{eq:a2}) is written as
$R_{\gen}^{(1,2)}=F^{(1,2)}(R_{\gen-1}^{(1)},R_{\gen-1}^{(2)},p)$. The three-variable function
$F^{(i)}(r_{1},r_{2},p)$ is the expected value of the inter-root-node resistance of
$G_{\text{D}}^{(i)}$ in which each edge has the resistance $r_{1}$ with probability $p$ or $r_{2}$
with probability $1-p$. By considering possible combinations of $r_{1}$ and $r_{2}$ for
$G_{\text{D}}^{(1)}$, we obtain
\begin{widetext}
\begin{align}
F^{(1)}(r_{1},r_{2},p)
=&p^{4}           \left(\frac{1}{2r_{1}}      +\frac{1}{2r_{1}}      \right)^{-1}+
  (1-p)^{4}       \left(\frac{1}{2r_{2}}      +\frac{1}{2r_{2}}      \right)^{-1}
 +4p^{3}(1-p)     \left(\frac{1}{r_{1}+r_{2}} +\frac{1}{2r_{1}}      \right)^{-1} \nonumber \\
&+4p(1-p)^{3}     \left(\frac{1}{r_{1}+r_{2}} +\frac{1}{2r_{2}}      \right)^{-1}
 +2p^{2}(1-p)^{2} \left(\frac{1}{2r_{1}}      +\frac{1}{2r_{2}}      \right)^{-1} \nonumber \\
&+4p^{2}(1-p)^{2} \left(\frac{1}{r_{1}+r_{2}} +\frac{1}{r_{1}+r_{2}} \right)^{-1} .
\label{eq:a4}
\end{align}
Similarly, $F^{(2)}(r_{1},r_{2},p)$ is given by
\begin{align}
F^{(2)}(r_{1},r_{2},p)
=&p^{6}            \left(\frac{1}{3r_{1}}      +\frac{1}{3r_{1}}      \right)^{-1}+
  (1-p)^{6}        \left(\frac{1}{3r_{2}}      +\frac{1}{3r_{2}}      \right)^{-1}
 +6p^{5}(1-p)      \left(\frac{1}{2r_{1}+r_{2}}+\frac{1}{3r_{1}}      \right)^{-1} \nonumber \\
&+6p(1-p)^{5}      \left(\frac{1}{r_{1}+2r_{2}}+\frac{1}{3r_{2}}      \right)^{-1}
 +6p^{4}(1-p)^{2}  \left(\frac{1}{r_{1}+2r_{2}}+\frac{1}{3r_{1}}      \right)^{-1} \nonumber \\
&+9p^{4}(1-p)^{2}  \left(\frac{1}{2r_{1}+r_{2}}+\frac{1}{2r_{1}+r_{2}}\right)^{-1}
 +6p^{2}(1-p)^{4}  \left(\frac{1}{2r_{1}+r_{2}}+\frac{1}{3r_{2}}      \right)^{-1} \nonumber \\
&+9p^{2}(1-p)^{4}  \left(\frac{1}{r_{1}+2r_{2}}+\frac{1}{r_{1}+2r_{2}}\right)^{-1}
 +2p^{3}(1-p)^{3}  \left(\frac{1}{3r_{1}}      +\frac{1}{3r_{2}}      \right)^{-1} \nonumber \\
&+18p^{3}(1-p)^{3} \left(\frac{1}{r_{1}+2r_{2}}+\frac{1}{2r_{1}+r_{2}}\right)^{-1} .
\label{eq:a5}
\end{align}
\end{widetext}
The resistances $R_{1}^{(1)}$ and $R_{1}^{(2)}$ are nothing but the inter-root-node resistances
$\Rroot^{(1)}$ and $\Rroot^{(2)}$ of $G_{\text{D}}^{(1)}$ and $G_{\text{D}}^{(2)}$, respectively.
These inter-root-node resistances are obviously $\Rroot^{(1)}=1$ and $\Rroot^{(2)}=3/2$. Thus,
$R_{2}^{(1)}$ and $R_{2}^{(2)}$ are computed as $R_{2}^{(1)}=F^{(1)}(1,3/2,1/2)=19601/15840$ and
$R_{2}^{(2)}=F^{(2)}(1,3/2,1/2)=14751747/7920640$. We then calculate $R_{\gen}^{(1,2)}$ iteratively
from Eq.~(\ref{eq:a2}) and the resistance $R_{\gen}$ is obtained by Eq.~(\ref{eq:a1}), namely
$R_{\gen}=(R_{\gen}^{(1)}+R_{\gen}^{(2)})/2$. The ratio $R_{\gen}/R_{\gen-1}$ converges rapidly to
$\Rrooteff=1.2398$ for $\gen\to \infty$.


\end{document}